\documentstyle[aps,prl,preprint]{revtex}
\def\Journal#1#2#3#4{{#1} {\bf #2}, #3 (#4)}

\def\NPB{{\em Nucl. Phys.} B}
\def\PLB{{\em Phys. Lett.}  B}
\def\PRL{\em Phys. Rev. Lett.}
\def\PRD{{\em Phys. Rev.} D}
\def\PR{\em Phys. Rev.}

\def\MPLA{{\em Mod. Phys. Lett.} A}
\def\PR{\em Phys. Rep.}
\def\PTP{\em Prog. Theo. Phys.}

\def\be{\begin{equation}}
\def\ee{\end{equation}}
\def\bea{\begin{eqnarray}}
\def\eea{\end{eqnarray}}

\draft
\begin{document}
\preprint{\vbox{
\hbox{SNUTP 96-030}\hbox{\tt hep-th/9604037}}}
\title{
COLLECTIVE COORDINATE QUANTIZATION OF DIRICHLET BRANES\footnote{
\tt Talk presented at the Workshop `Frontiers in Quantum Field Theory' in 
honor of 60th birthday of Prof. K. Kikkawa, Osaka, Japan, December 14-17,1995.
To appear in the Proceedings.} }
\author{ SOO-JONG REY}
\address{Physics Department, Seoul National University, Seoul 151-742 KOREA}

\maketitle
\begin{abstract}
Collective coordinate quantization of Dirichlet branes is discussed.
Utilizing Polchinski's combinatoric rule, semiclassical D-brane wave 
functional is given in proper-time formalism. D-brane equation of motion 
is then identified with renormalization group equation of defining Dirichlet 
open string theory. Quantum mechanical size of macroscopically charged D-brane 
is illustrated and striking difference from ordinary field theory BPS particle 
is emphasized.
\end{abstract}
\date{}
\section{Introduction}

Recently nonperturbative string theory has given us
many surprising results. There are now compiling evidences that all
known perturbatively defined string theories are related each other
by duality at nonperturbative level~\cite{hulltownsend}. 
Central to this advance was progress to semiclassical string theory, 
in particular, deeper understanding of stringy topological solitons 
over the last few years~\cite{dh,mine8991,mine91,st,chs,duff}. 
Previous study of string solitons, however, has been restricted mainly to
low-energy effective field theory approximation. While exact conformal 
field theories in a few cases has been known from the earliest 
days~\cite{mine8991,chsrev}, further progress was hampered because 
of technical difficulties in dealing with geodesic motion interpolating 
between different conformal field theories, viz. space of nontrivial
vacua of string field theory. 

In a recent remarkable work~\cite{pol}, Polchinski has obtained 
an exact conformal field theory describing Ramond-Ramond charged solitons. 
These so-called D(irichlet)-branes are described in terms of Dirichlet 
open strings that are coupled to the underlying type II closed strings.
Polchinski's work has cleared up many puzzling aspects that arose previously
when string solitons were studied within low-energy effective field theory
truncation. The D-brane proposal has already passed many nontrivial 
consistency tests but all of them so far were mainly on static properties.
With its simplicity and exactness it should now be possible to study 
\sl quantum dynamics \rm of string solitons in detail. 

In this talk I report my recent work~\cite{mine} on several aspects of 
D-brane dynamics: 
collective coordinates, semi-classical quantization, renormalization group 
interpretation of equation of motion and quantum mechanical size of 
macroscopically charged D-brane.

\section{D-Brane and Collective Coordinates}
Consider a conformally invariant two-dimensional system. If a boundary
is introduced at which the bulk system ends, then it is well-known that
microscopic detail is renormalized into a set of conformally invariant
boundary conditions. For a Gaussian model such boundary conditions are
either Neumann or Dirichlet boundary conditions but not a combination 
of the two. 

Similarly, in closed string theory, it is possible to introduce 
worldsheet boundaries. At each boundaries, string coordinates
$X^\mu(z, {\overline z})$ may be assigned to either Neumann (N) or 
Dirichlet (D) boundary conditions. Mixed boundary condition may seem
break 10- or 26-dimensional Lorentz invariance. However, on toroidally 
compactified spacetime, target space duality $R \leftrightarrow \alpha'/R$ 
interchanges N and D boundary conditions. Hence Lorentz invariance is
maintained up to target space $T$-duality. Denote N-coordinates as 
$X^i, i = 0, 1, \cdots, p$ and D-coordinates as $Y^a, a=p+1, \cdots, 9 (25)$. 
Each worldsheet boundaries are mapped into spacetime \sl extrinsic \rm 
hypersurfaces of dimension $(p+1)$ spanned by $X^i$, viz. Dirichlet p-brane 
world-volume.  
Polchinski~\cite{pol} has shown these
D-branes are nonperturbative states of type II strings that carry
RR-charges obeying Dirac quantization condition and that saturate
BPS bounds.

Worldsheet chiral symmetries restrict possible p-branes further. 
Type IIB strings are worldsheet symmetric that even numbers of 
D-coordinates are possible, hence, contains $p=-1,1,3$ branes and 
their magnetic duals.
Similarly, for type IIA odd numbers of D-coordinates are allowed, 
viz. $p=0,2,4$ branes and their magnetic duals.
Since IIA and IIB string theories are mapped into each other under
target space duality $R \rightarrow \alpha' /R$, one can build up all D-branes
from the oriented open string sector ($p=9$) in IIB theory and
cascade $T$-duality transformations.
In type I string theory, because of worldsheet orbifold projection,
only $p=1,5,9$ branes are allowed. 

Worldsheet interaction of type II strings with D-branes are described by 
Dirichlet open string theory. Worldsheet interaction at each boundaries
is deduced by cascade $T$-duality transformation of the known oriented
$9$-brane (open string) theory. For massless excitations, the worldsheet
interactions at each boundaries are described by
\begin{equation}
S_B = \oint d \tau 
\Big[ \sum_{i=0}^p A_i (X^0, \cdots, X^p) \partial_t X^i
+ \sum_{a=p+1}^{9(25)} \phi_a (X^0, \cdots, X^p) \partial_n Y^a]
\end{equation}
Image of `Chan-Paton quark' (end of Dirichlet open string) is mapped 
onto D-brane world-volume $\Sigma_{p+1}$, but otherwise can move 
freely on it by $(p+1)$-dimensional translation invariance. 
Such restriction is consistent if gapless gauge field excitation 
described by the vertex operator $V_A = \oint A_i \partial_t X^i$ 
is present only on $\Sigma_{p+1}$ hypersurface but not outside,
hence, $A_i = A_i (X^j)$.
This world-volume gauge fields also mix with type II Kalb-Ramond field 
$B_{\mu \nu}$ as is evident from bulk-extended expression of the vertex 
operator $V_A = i \int d^2 z \partial_z (A_i \partial_{\overline z} X^i)
- (c.c.)$. This is the well-known Cremmer-Scherk~\cite{cremmerscherk} 
coupling and generates stringy Higgs mechanism.
The coupling also makes it clear what the meaning of world-volume gauge 
field is:
closed string winding modes transmutes into massive world-volume gauge 
fields and provides $SL(2,{\bf Z})$ orbits of Neveu-Schwarz and Ramond
charges to the D-brane BPS mass.

The vertex operator $V_\phi = \oint \phi_a (X^i) \partial_n Y^a$ describes 
transverse translation of local D-brane world-volume $X^i$, hence,
\sl collective coordinates \rm. Normally spacetime translations are 
redundant and correspond to null states. This is clear from rewriting
$V_\phi = \int d^2 z \partial_z ( \phi_a \partial_{\overline z} Y^a) + 
(c.c.)$,
which decouples on a compact worldsheet. The decoupling fails precisely 
when D-boundaries are present and $V_a$'s turn into genuine physical modes.
This is consistent with spacetime point of view since, in the presence
of a D-brane, translational symmetry is spontaneously broken and new
Goldstone mode states should appear. The $V_a$'s that fail to decouple
and fail to be null are precisely those Goldstone states.

Low-energy spacetime interactions type II strings with $N$ independent 
D-branes are then described by massless modes of Dirichlet string theory: 
$D=10, N=2B$ supergravity coupled to $D=10, N=1$ Dirac-Born-Infeld 
$U(N)$ gauge theory on $\Sigma_{p+1}$ dimensionally reduced and $T$-dualized
onto $\Sigma_{p+1}$. Mismatch of spacetime supersymmetry and $U(N)$ 
Chan-Paton gauge group does not cause problems since the D-brane excitations 
are confined only on $\Sigma_{p+1}$. Thus, dimensionally reducing and
$T$-dualizing first and then making a `nonrelativistic expansion' for
small gauge and Goldstone field excitations~\cite{witten}, 
\begin{eqnarray}
S_{\rm wv} &=& - {\rm Tr} T_p \int_{\Sigma_{p+1}} \!\! e^{-\phi} 
\sqrt { {\rm det} ( G_{MN} + B_{MN} + F_{MN}) }
\nonumber \\
&\rightarrow& {\rm Tr} T_p \int_{\Sigma_{p+1}} \!\! e^{-\phi} 
{\sqrt G} \Big( 1+ {1 \over 4} (F+B)_{ij}^2 + [D_i, \phi^a]^2 + 
[\phi^a, \phi^b]^2 \Big).
\end{eqnarray}
Here $T_p e^{-\phi}$ denotes $p$-brane static mass density obtained from T-dual 
transform of 9-brane dilaton tadpole amplitude on a disk~\cite{tadpole}.
The first and second 
terms give bare energy and Casimir energy of static D-branes. The second 
term also contains aforementioned Cremmer-Scherk coupling, hence, 
manifest gauge invariance of Kalb-Ramond $B_{\mu \nu}$ field is maintained. 
The third and fourth terms are kinetic and potential energy for moving 
D-branes.

\section{Combinatorics of Perturbative Dirichlet String Theory}
Consider string S-matrix elements involving D-branes. Dirichlet string 
theory associates each D-brane to a Dirichlet boundary on which lives 
an independent `Chan-Paton quark'.  Interaction between D-branes and 
string states are then described by Riemann surfaces with handles and holes.
This also implies that apparently disconnected worldsheet diagrams are 
in fact connected in spacetime as long as boundaries from disconnected
worldsheet are mapped into common D-brane(s).
This entails a new rule of string perturbation expansions for the S-matrix
generating functional $Z$ ~\cite{polcom}. 
Let $n=1, \cdots, N$ label $N$ independent D-branes into which different 
species of `Chan-Paton quarks' are mapped. 
Worldsheet perturbation theory for $Z$ is then organized as 
\begin{equation}
Z_{\rm new} = \sum_{N=0}^\infty {1 \over N!} \otimes \Big(\prod_{n=1}^N 
\int [d Y^a_n] \Big) \otimes \sum_{h = 0}^\infty {1 \over h!} \otimes 
\sum_{a_1, \cdots, a_h = 1}^N {h! \over m_1! m_2! \cdots m_N!}
\end{equation}
where $m_i \ge 0, \sum_i m_i = h$.
For fixed $N$, we sum over the number $h$ of worldsheet boundaries 
and sum each of the $h$ `Chan-Paton quarks' (independent D-branes) from 1 
to $N$. Then we integrate over the transverse positions of each D-branes
and finally sum over the total number $N$ of D-branes. Since summing over
the number of holes and the `Chan-Paton quarks' amounts to summing over the
number of holes mapped into a given D-brane, the above combinatorics
can be organized as 
\begin{eqnarray}
Z_{\rm new} &=& \sum_{N=0}^\infty \otimes {1 \over N!} 
\int \prod_{n=1}^N [dY^a_n] \otimes \sum_{h_1=0}^\infty
{1 \over h_1!} \sum_{h_2=0}^\infty {1 \over h_2!} \cdots  
\sum_{h_N=0}^\infty {1 \over h_N!}
\nonumber \\
&=&
\sum_{N=0}^\infty \otimes {1 \over N!} \prod_{n=1}^N \Big(\int [dY^a_n]
\sum_{h_n=0}^\infty {1 \over h_n!} \Big)
\end{eqnarray}
Exponentiation of disconnected worldsheets then generate a complete
S-matrix generating functional $Z$.

In old perturbation theory~\cite{green} each Dirichlet boundaries are
mapped into independent spacetime points, subsequently integrated over 
\begin{equation}
Z_{\rm old} 
= \sum_{N=0}^\infty {1 \over N!} \prod_{n=1}^N \int [d Y^a_n]
\sum_{h_n=0}^\infty {1 \over n_a!} \delta_{n_a, 1}
\end{equation}
The difference arises because the D-branes are extrinsic structure
to spacetime.

Combinatorics for the S-matrix generating functional in Dirichlet string 
perturbation theory may be rephrased as follows. 
Prepare for $m$ disconnected, compact Riemann surfaces, 
create $n_h$ holes arbitrarily distributed among the $m$ Riemann surfaces.
Map each holes to the world-volume of $N$ independent D-branes
allowing duplications. Finally sum over $m, n_h, N$ independently with
appropriate combinatoric factors ${\cal S}_{\rm ms}$ of $n_h$ boundaries 
into $m$ Riemann surfaces and ${\cal S}_{\rm st}$ of $n_h$ boundaries into 
$N$ D-branes 
\begin{equation}
Z_{\rm new} = \sum_{n_h=0}^\infty {1 \over n_h!}
\otimes
\Big(
\sum_{N=0}^\infty {1 \over N!} {\cal S}_{\rm st} (N \leftarrow n_h)
\otimes \sum_{m=0}^\infty {1 \over m!} {\cal S}_{\rm ws} (n_h \rightarrow m)
\Big).
\end{equation}
We note that a single exponentiation maps each Dirichlet boundaries 
doubly into disconnected Riemann surfaces and into independent D-brane 
world-volumes in a symmetric manner.

\section{Semi-Classical Wave Function of D-Branes}
Consider Dirichlet string partition function in the background of 
the type II string fields in which all worldsheet boundaries are mapped 
into a single
D-brane world-volume $\Sigma_{p+1}$. The partition function serves as a 
generating functional, hence, S-matrix elements between D-brane and string 
states are derived from local variation of background string fields. 
The partition function is also related to the (Euclidean) wave functional 
$\Psi_1$ of the D-brane. The new combinatoric rule relates the wave 
functional to the partition function 
\begin{equation}
Z_1 = \int [dY^a]\Psi_1 [Y^a(\cdot)] \hskip1cm \Psi_1[Y^a] = e^{{\cal S}_1}.
\end{equation}
Here ${\cal S}_1$ sums up all one-particle irreducible 
\sl connected \rm worldsheet diagrams, 
whose boundaries are mapped to the D-brane world-volume.
Integration over world-volume gauge field is already made for $\Psi_1$ 
to ensure type II winding quantum number conservation.
Dirichlet string perturbation theory yields
\begin{equation}
{\cal S}_1[Y^a] = \sum_{h=1}^\infty e^{\phi(h-2)} S_{(h)}
\end{equation}
in which $S_{(h)}$ denotes amplitude with $h$-holes. Sum over handles 
is implicitly assumed in the definition of $S_{(h)}$.

Higher order contributions $S_{(h \ge 2)}$ come from annulus, torus
with a hole etc or sphere with three holes etc. They amount to 
D-brane mass renormalization. For type II string all except the disk 
diagram ($h=1$) vanishes identically because of spacetime supersymmetry
nonrenormalization theorem. For D-instanton, this is consistent with known 
results that the RR instantons are exact to all orders in string 
perturbation theory.
The disk amplitude $S_{(1)}$ for type II superstring is easily obtained 
from 9-brane boundary state~\cite{clny} after appropriate $T$-duality
transformations. For simplicity, keeping only the transverse fluctuation 
of the D-brane world-volume
\begin{eqnarray}
{\cal S}_1 &=& T_p \int d \Sigma_{p+1} e^{-\phi} 
{\sqrt {{\rm det} G_{ij}}}
\nonumber \\
&=& \int d \Sigma_{p+1} \Big[ {1 \over V} {\rm det} G_{ij}
- M^2_p V \Big]
\end{eqnarray}
where $M_p = T_p e^{-\phi}$. In the last expression, we have also introduced 
non-dynamical `proper-time' variable $V$. Functional integral over $V$ 
introduces no new Jacobian and amounts to a sum over all possible 
propagation of D-brane.
 
Consistency of Dirichlet string coupled to type II string requires 
to maintin conformal invariance or BRST invariance. The wave functional, 
however, contains various sources of logarithmic divergences that 
violate the conformal or BRST invariance.
We have seen earlier that the Goldstone mode vertex operators are \sl 
isolated \rm, descendent operators that fail to decouple and to be null 
in the presence of the Dirichlet boundaries. It is now necessary to 
examine carefully all possible boundaries of moduli space.
Crucial understanding on how to do this has been made in a recent
important work by Fischler \sl et.al.\rm~\cite{fischler}.
Consider a finite but large proper-time interval $T$
so that all scattering states of the D-brane form a discrete set 
of $L_0$ and ${\overline L}_0$ spectrum separated by a gap from the
continuum of type II string states. With this infrared regularization
provided we can properly extract divergent amplitudes unambiguously.

Consider the disk amplitude near a boundaries of moduli space for two 
sets of closed string vertex operators.
Inserting a complete set of states labelled by $\{a\}$ that includes
those naively BRST null and denoting propagators as $\Pi_a$,
the disk amplitude
\begin{eqnarray}
\langle \cdots \rangle_{D_2} &\rightarrow& 
\int \hskip-0.6cm \sum_{\rm states} 
\, \langle \cdots | a \rangle_{D_2} \Pi_a(k) \langle a | \cdots \rangle_{D_2}
\nonumber \\
&=& \int \! {d^D k \over (2 \pi)^D} \, \langle \cdots V_A (k) \rangle_{D_2}
\langle V^\dagger_A (k) \cdots \rangle_{D_2} 
\times \int_0^\infty {d t \over t} 
e^{-t (k_n^2 + m_n^2)}
\nonumber \\
&+& \sum_{a=p+1}^{D-1} \langle \cdots V_\phi \rangle_{D_2} 
\langle V_\phi^\dagger \cdots \rangle_{D_2}
\times \int_\epsilon^\infty {dt \over t} .
\end{eqnarray}
The first comes from physical excitations with continuum distribution 
labelled by momenta $k_a$, hence, 
does not cause any infrared divergence. The second is due to 
intermediate exchange of the D-brane Goldstone mode. 
Spacetime picture is that a tiny Dirichlet open string state propagates for
a long proper-time and diverges linearly. 
As the D-brane Goldstone mode spectrum 
is discrete and isolated for a finite proper-time interval $T$, 
it is not possible to analytically continue kinematics and avoid infrared 
divergence as the cutoff $\epsilon \rightarrow 0$. It is precisely these 
divergences we need to cure.

Similarly the annulus amplitude near a boundary of moduli space at which
the annulus strip is pinched into a thin, long open string propagation.
While vanishing for stationary D-brane (BPS static force balance 
condition), the annulus amplitude with time varying D-brane velocity 
and/or  with a background to soak up all the spacetime fermion zero modes
are nonvanishing. Such amplitude also contains divergences
\begin{eqnarray}
\langle \cdots \rangle_{A_2} &\rightarrow &
\int \hskip-0.6cm \sum_{\rm states} \langle a | \cdots |a \rangle
\Pi_a(k)
\nonumber \\
&=& \int {d^D k \over (2 \pi)^D} \langle \cdots V_A (k) V_A^\dagger (k) 
\cdots \rangle_{D_2}
\times  \int_0^\infty {dt \over t} e^{-t(k_n^2 + m_n^2)} 
\nonumber \\
&+& \sum_{a=p+1}^{D-1} \langle \cdots V_{\phi_a} V_{\phi_a}^\dagger
\cdots \rangle_{D_2} \times  \int_\epsilon^\infty {dt \over t} .
\end{eqnarray}
This diagram contains also infrared divergence due to D-brane
Goldstone-mode exchange.
Again spacetime picture is that a tiny Dirichlet open string propagates
for a long proper-time interval and diverges linearly.

Noting that $V_{\phi_a} = \oint \phi_a \partial_n Y^a = 
\phi_a \partial / \partial Y^a$
viz. rigid translation of D-brane by $\phi_a$ transversally
we find the two logarithmically divergent contributions combine into 
a total derivative
\begin{equation}
\big(\langle .. \rangle_{D_2} + \langle .. \rangle_{A_2} 
\big)_{\log \epsilon} = (\log \epsilon) \Big[ 
{1 \over 2!} \phi \cdot \nabla_Y \langle .. \rangle_{D_2} 
\phi \cdot \nabla_Y \langle .. \rangle_{D_2} 
+ \langle .. \rangle_{D_2} (\phi \cdot \nabla_Y)^2
\langle .. \rangle_{D_2} 
\Big] ,
\end{equation}
hence, to this order in $e^\phi$,
D-brane wave functional $\Psi_1[Y]$ contains
\begin{equation}
(\exp[ \langle .. \rangle_{D_2} + \langle .. \rangle_{A_2} ])_{\log \epsilon}
\approx (\log \epsilon) \, {1 \over 2!} (\phi \cdot \nabla_Y)^2 
 \exp[{\langle \cdots \rangle_{D_2}} + \langle .. \rangle_{A_2}] 
 .
\end{equation}
We have isolated leading $\log \epsilon$ divergences due to 
\sl worldsheet \rm short-distance singularity in the presence of 
Dirichlet boundaries.
In \sl spacetime, \rm the divergence arises from propagation of
isolated D-brane collective coordinate modes as the proper-time
interval $T \rightarrow \infty$. Hence, the two regulators 
may be identified as $T \approx -\log \epsilon$ up to multiplicative 
and additive constants that can be determined from explicit S-matrix 
element calculations~\cite{mine}.
Logarithmic relation between the two should be evident if 
one recalls proper-time formalism~\cite{feynman} of Polyakov path integral: 
dilatation of worldsheet coordinates $\epsilon \rightarrow e^l \epsilon$ 
corresponds to shift of proper-time $T \rightarrow T - \log l$.

\section{D-Brane Equation of Motion and Renormalization Group Flow}
Having isolated divergences in the presence of D-brane, how do we cure of 
them? Governing principle of string theory is the requirement of conformal
or BRST invariance. Much the way spacetime equations of motion of string 
background fields has been obtained, the requirement applied to the
Dirichlet string theory is expected to a new equation for consistent 
D-brane dynamics.  With this motivation we now require 
\begin{equation}
\epsilon {d \over d \epsilon} Z_1 = \epsilon {d \over d \epsilon}
\Big( \int [dY^a] \Psi_1[Y^a] \Big) = 0.
\end{equation}
Since $Z_1$ and $\Psi_1$ sums up worldsheet diagrams of arbitrary
number of handles and holes, Eq.(14) invokes Fischler-Susskind~\cite{fs} 
mechanism in an essential way.

There are two possible ways to achieve this requirement.
Noting that Eq.(13) is a total derivative with respect to the zero modes 
$Y^a$'s, the first is to require that \sl integral \rm of $\Psi_1$  
, viz. $Z_1$ itself satisfies conformal invariance requirement. 
This viewpoint has been advocated by Polchinski~\cite{pol}: 
logarithmically divergent part Eq.(13) is a total derivative in $Y^a$-space
and drops out upon integration over $Y^a$'s in $Z_1$.
Obviously in cases we are interested in \sl local \rm
dynamics of D-brane this requirement does not offer 
much information. For example, given the semi-classical wave functional 
$\Psi[Y^a]$ first, how do we uncover an equation of motion to which the 
wave functional satisfies?
The second viewpoint is then that the \sl integrand \rm of the path
integral $Z_1$, viz. $\Psi_1$ is free from infrared divergence.
Adopting this view point we get 
\begin{equation}
\epsilon {d \over d \epsilon} \Psi_1 [Y^a]
= {\phi^2 \over 2!} (\nabla_Y)^2 \Psi_1 [Y^a].
\end{equation}
Recalling that worldsheet variable $\log \epsilon$ is linearly to 
the spacetime proper-time interval $T$, the equation looks strikingly
simiar to the Euclideanized Schr\"odinger equation. To show that this 
is not a mere coincidence, let us go back to the procedure of isolating
logarithmic divergences in disk and annulus amplitudes. 
Conformal invariance requirement to $\Psi_1[Y]$ amounts to  
Wilson renormalization group equation for Dirichlet boundary action 
\begin{equation}
\epsilon {\partial \over \partial \epsilon} \Psi_1 [Y] = {1 \over 2!} 
\oint d \tau_1 \oint d \tau_2  \epsilon \partial_\epsilon G_{ab}(t_1, t_2) 
{\partial \over \partial Y^a(t_1)} {\partial \over \partial Y^b (t_2)}
\Psi_1 [Y] .
\end{equation}
An important point is that the Dirichlet boundary Green function 
$G_{ab}(t)$ for transverse coordinates contains zero-mode part
\begin{equation}
G^{ab}(t_1,t_2) = \langle : Y^a (t_1) Y^b (t_2): \rangle + (- \log \epsilon)
|{\overline Y}^a|^2 \delta^{ab}.
\end{equation}
The zero-mode ${\overline Y}^a$ of the transverse coordinates is a direct 
reflection of the spacetime zero-modes associated with the D-brane recoil.
The zero mode is independent of the Dirichlet string worldsheet variables 
and is proportional to $\log \epsilon$. 
Earlier identification that $-\log \epsilon \approx T$ 
also supports the interpretaion. The zero-mode is precisely the new 
source of conformal
and BRST anomalies we have explicitly isolated in the previous section
Eqs.(11) - (13).

So far we have examined the single logarithmic divergences and ways of
ensuring their cancellation. Multiple logarithms are similarly cancelled
as has been explicitly shown up to double logarithms for
9-branes~\cite{dasrey}. The leading logarithms may be resummed and
exponentiated to a new wave functional
\begin{equation}
\Psi_V [Y^a]
= e^{(-\log \epsilon) {\nabla_Y^2 \over 2 M_p} } \Psi[Y^a].
\end{equation}
Physical meaning of this is as follows: recoiling D-brane acquires
transverse kinetic energy $P^2 / 2 M_p = M_p V^2 / 2$.
During time interval $T \approx - \log \epsilon$, the (Euclidean)
wave function acquires an additional phase (action) proportional to the
kinetic energy. The leading log resummation is necessary
since the kinetic energy is of the same order as the static energy
${\cal O}(1/\lambda)$ even though suppressed by velocity-squared.
Since the new wave functional $\Psi_V$ describes consistently a
boosted D-brane, conformal or BRST invariance implies 
\begin{equation}
\Big( {d \over d T} + { {\nabla}_Y^2 \over 2 M_p} \Big) \Psi_V [Y^a] = 0.
\end{equation}

Wilson renormalization group equation has been previously 
proposed~\cite{rge} as a defining principle for obtaining string 
field equations of motion.
The idea hs been extended to take into account of the Fischler-Susskind 
mechanism~\cite{das}. When applied to Dirichlet string theory we now 
see that consistent \sl D-brane equation of motion \rm Eq.(19)
follows from the renormalization group equation.
Equivalently, the equation can be understood as a consequence of 
on-shell Ward identity of type II string in BRST formulation~\cite{brst}. 
Type II string 
contains BRST invariant conserved charges associated with translational 
invariance. The Dirichlet string boundary action Eq.(1) added to the 
type II string is naively BRST exact perturbations but fails precisely
in the presence of D-brane. This means that the boundary action is a total 
BRST derivative of `bad' operators that fail to decouple. 
Ward identities of spontaneously broken translational symmetry 
is then realized through non-decoupling of these `bad' opearators 
and gives rise to a `quantum master equation' similar to Eq.(15).

Full consideration of D-brane dynamics 
may require more careful analysis of dynamical gravity effect on the 
embedded D-brane world-volume. 
Previous experience with noncritical string theory~\cite{daswadia,susskind} 
hints renormalization group flow equation changes time derivative 
in Eqs.(15,16,19) from first- to second-order 
\begin{eqnarray}
&& \big[\partial_T^2 - 2 M_p \partial_T - \nabla_Y^2 \big] \Psi_V [Y^a] = 0
\nonumber \\
&\rightarrow& \big[ \partial_T^2 - \nabla_Y^2 - M_p^2 \big]
e^{-M_p T} \Psi_V [Y^a] = 0,
\end{eqnarray}
viz., a covariant equation of motion for D-brane emerges.
Similarly, massive Dirichlet string exchange is expected to generate contact 
interactions among D-branes and gives rise to nonlinear equation of motion
\cite{susskind}. 

\section{Quantum Aspects of Macroscopically Charged D-Brane}
So far I have discussed exclusively one-body aspects of D-brane.
I now turn to a many-body aspects of macroscopically charged D-brane.
Low-energy excitation of $N$ overlapping D-branes is described 
by~\cite{witten} 
dimensionally reduced $D=10$ supersymmetric $U(N)$ Yang-Mills theory on 
$\Sigma_{p+1}$
\begin{equation}
S_{\rm wv} = {\rm Tr} T_p e^{-\phi} \int_{\Sigma_{p+1}} {\sqrt G}
[ {1 \over 4} F_{ij}^2  + (D_i \phi^a)^2 + \cdots]
\label{wvaction}
\end{equation}
Essential many body-aspects is already present for type IIA D-particles 
, so I consider this case first. For macroscopically charged D-particles 
$N \rightarrow \infty$, world-line action is $U(N)$ \sl matrix \rm 
supersymmetric quantum mechanics.
Gauge potential $A_0$ is nondynamical and but constrains the
D-particle displacement $\phi^a (t)$ to a gauge singlet configuration.
Diagonalizing the D-particle displacements $\phi^a$
\begin{equation}
\phi^a = U {\rm diag} (y^1(t), y^2(t), \cdots, y^N(t)) U^\dagger,
\end{equation}
low-energy excitation is governed by an effective Hamiltonian
\begin{equation}
H_{\rm D-BPS}
= {\rm Tr} \sum_{A=1}^8 \{ Q_A, Q^\dagger_A \}
= - {1 \over 2 M_0} \sum_{a=1}^N \nabla_a^2 
+ {1 \over 2} \sum_{a \ne b}^N \log(y_a - y_b)
+ \sum_{a \ne b}^N | y^a - y^b|
+ \cdots
\label{efflag}
\end{equation}
The second term is quantum effective potential coming from functional
integral after the diagonalization Eq.(22).

I now compare this with  an effective Hamiltonian of macroscopically
charged, field theory BPS particles.
For the simplest BPS particles such as kinks in one dimension, 
low-energy dynamics is described entirely in terms of position of 
each particles. Hence, locally in the $N$-particle moduli space, 
effective action is given by $N$-dimensional \sl vector \rm 
supersymmetric quantum mechanics
\begin{equation}
H_{\rm BPS} = {1 \over 2} \sum_A \{Q_A, Q^\dagger_A \}
= -{1 \over 2} \nabla^2 +
{1 \over 2}  (\nabla W)^2 + {1 \over 2}
{\partial^2 W \over \partial x^i \partial x^j}
[\psi^{i \dagger}, \psi^j]
\end{equation}
Because of mutual force balance between BPS particles 
$W \approx 0$, hence, of ideal gas type. 

It is now clear how macroscopically charged D-particle behave 
differently from field theory BPS-particle.
At classical level D-particles behave indifferently from BPS
particles: both experience no net force because of BPS nature. 
At quantum level, however, D-particles experience logarithmically
repulsive quantum effective potential. Because of this quantum 
pressure, average spacing between constituent D-particles
is of string scale ${\cal O}(\sqrt {\alpha'})$.
Collective excitation of D-particle gas is that of one-dimensional
Bose gas described by two-dimensional free scalar field theory.
For field theory D-particles, no quantum effective potential
, hence, no quantum pressure is generated. So long as intrinsic size 
is ignored these BPS particles can overlap freely. 

The above argument is not restricted to D-particles but extends to
other D-branes. For instance, consider 8-branes in type IIB superstring
theory. If compactified on a circle and worldsheet orbifoldized, 
one obtains type-$I'$ string. Uniformly weak coupling configuration is
when two sets of 16 8-branes are located at the ${\bf Z}_2$ fixed
points $X^9 = 0, 2 \pi / R_I$. While 16 
is not a terribly large number, let us pretend so and study many-body
aspects. Their low-energy excitation is given by the transverse
locations of 8-brane center of masses. Again this is described by
$N=8$ supersymmetric $O(16)$ \sl matrix \rm quantum mechanics projected 
to a gauge singlet sector. At quantum level, equilibrium positions are 
when the inter-spacing of elementary 8-branes is of order $\sqrt {\alpha'}$.

\section*{Acknowledgments} 
I am grateful to Professor Kikkawa for warm hospitality and Professors
H. Itoyama, M. Ninomiya for organizing an enjoyable conference.
This work was supported in part by U.S.NSF-KOSEF Bilateral Grant,
KRF Nondirected Research Grant and International Collaboration Grant,
KOSEF Purpose-Oriented Research Grant and SRC-Program,
and Ministry of Education BSRI 94-2418.



\begin{thebibliography}{99}
\bibitem{hulltownsend} C. Hull and P. Townsend, 
\Journal{\NPB}{438}{109}{1995};
E. Witten, \Journal{\NPB}{443}{85}{1995}.

\bibitem{dh} A. Dabholkar and J.A. Harvey, \Journal{\PRL}
{63}{478}{1989}.

\bibitem{mine8991} S.-J. Rey, \sl Axionic String Instantons \& Their
Low-Energy Implications, \rm in {\em Superstring and Particle Theory},
eds. L Clavelli, B Harms, p.291-300 (World Scientific, Singapore, 1990);
\sl On String Theory and Axionic Strings \& Instantons, \rm
in {\em Particles and Fields '91}, eds. D Axon, D Bryman, M Comyn,
p.876-881 (World Scientific, Singapore, 1992).

\bibitem{mine91} S.-J. Rey, \Journal{\PRD}{43}{526}{1991}.

\bibitem{st} A. Strominger, \Journal{\NPB}{343}{167}{1990},
Erratum \Journal{}{353}{565}{1991}.

\bibitem{chs}
C.G. Callan, J.A. Harvey and A. Strominger,
\Journal{\NPB}{359}{611}{1991}; \Journal{\NPB}{367}{60}{1991}.

\bibitem{duff}
For a recent review, see M.J. Duff, R.R. Khuri and J.X. Lu, 
\Journal{\PR}{259}{213}{1995}.

\bibitem{chsrev} C.G. Callan, J.A. Harvey and A. Strominger, 
\sl Supersymmetric String Solitons, \rm in {\em String Theory and 
Quantum Gravity '91}, eds.  , 
p.208-244 (World Scientific, Singapore, 1992).

\bibitem{pol} J. Polchinski, \Journal{\PRL}{75}{4724}{1996}.

\bibitem{mine} S.-J. Rey, {\em Collective Coordinate Dynamics and
Semiclassical Quantization of Dirichlet Brane,}
SNUTP 96/014 preprint (1996).

\bibitem{cremmerscherk} E. Cremmer and J. Scherk, \Journal{\NPB}
{50}{222}{1972}.

\bibitem{witten} E. Witten, \Journal{\NPB}{460}{335}{1996}.

\bibitem{tadpole} S.-J. Rey, \Journal{\PLB}{203}{393}{1988};
J. Liu and J. Polchinksi, \Journal{\PLB}{203}{39}{1988}.

\bibitem{polcom} J. Polchinski, \Journal{\PRD}{50}{R6041}{1994}.

\bibitem{green} M.B. Green, \Journal{\PLB}{69}{89}{1977}; \Journal
{\PLB}{201}{42}{1988}; \Journal{\PLB}{282}{380}{1992}; M.B. Green
and J. Polchinski, \Journal{\PLB}{335}{377}{1994}.

\bibitem{clny} C.G. Callan, C. Lovelace, C.R. Nappi and S.A. Yost,
               \Journal{\NPB}{308}{221}{1988}.

\bibitem{fischler} W. Fischler, S. Paban and M. Rozali,
\Journal{\PLB}{352}{298}{1995}.

\bibitem{feynman} R.P. Feynman, \Journal{\PR}{80}{440}{1950};
Y. Nambu, \Journal{\PTP}{5}{82}{1950}.
See also B. Sathiapalan, \Journal{\NPB}{294}{747}{1987}.

\bibitem{fs} W. Fischler and L. Susskind, \Journal{\PLB}{171}{383}{1986};
             \Journal{\PLB}{173}{262}{1986}.

\bibitem{dasrey} S.R. Das and S.-J. Rey, \Journal{\PLB}{186}{328}{1987}.

\bibitem{rge} C. Lovelace, \Journal{\NPB}{273}{413}{1986};
T. Bank and E. Martinec, \Journal{\NPB}{294}{733}{1987};
J. Hughes, J. Liu and J. Polchinski, \Journal
	      {\NPB}{316}{15}{1989}.

\bibitem{das} S.R. Das,  \Journal{\PRD}{38}{3105}{1988}.

\bibitem{brst} E. Witten, \Journal{\PRD}{46}{5467}{1992};
E. Verlinde, \Journal{\NPB}{377}{141}{1992}.

\bibitem{daswadia} S.R. Das, S. Naik and S.R. Wadia,
                   \Journal{\MPLA}{4}{1033}{1989}.

\bibitem{susskind} A. Cooper, L. Susskind and L. Thorlacius,
                   \Journal{\NPB}{363}{132}{1991}.

\end{thebibliography}
\end{document}